\documentclass[pre,aps,twocolumn,,amsmath,amssymb,showpacs,showkeys]{revtex4-1}

\usepackage{amsmath}
\usepackage{amstext}
\usepackage{amsfonts}
\usepackage{amssymb,amscd}
\usepackage{amsopn}
\usepackage{graphicx}
\usepackage{xcolor}
\usepackage{bm,bbm}
\usepackage{epsfig,epic}
\newtheorem{theorem}{Theorem}
\newtheorem{lemma}[theorem]{Lemma}

\newtheorem{proposition}[theorem]{Proposition}
\newenvironment{proof}[1][Proof]{\noindent\textbf{#1.} }{\ \rule{0.5em}{0.5em}}

\newcommand{\newc}{\newcommand}
\newc{\beq}{\begin{equation}}
\newc{\eeq}{\end{equation}}
\newc{\kt}{\rangle}
\newc{\br}{\langle}
\newc{\beqa}{\begin{eqnarray}}
\newc{\eeqa}{\end{eqnarray}}
\newc{\pr}{\prime}
\newc{\longra}{\longrightarrow}
\newc{\ot}{\otimes}
\newc{\rarrow}{\rightarrow}
\newc{\h}{\hat}
\newc{\bom}{\boldmath}
\newc{\btd}{\bigtriangledown}
\newc{\al}{\alpha}
\newc{\be}{\beta}
\newc{\ld}{\lambda}
\newc{\sg}{\sigma}
\newc{\p}{\psi}
\newc{\eps}{\epsilon}
\newc{\om}{\omega}
\newc{\mb}{\mbox}
\newc{\tm}{\times}
\newc{\hu}{\hat{u}}
\newc{\hv}{\hat{v}}
\newc{\hk}{\hat{K}}
\newc{\ra}{\rightarrow}
\newc{\non}{\nonumber}
\newc{\ul}{\underline}
\newc{\hs}{\hspace}
\newc{\longla}{\longleftarrow}
\newc{\ts}{\textstyle}
\newc{\f}{\frac}
\newc{\df}{\dfrac}
\newc{\ovl}{\overline}
\newc{\bc}{\begin{center}}
\newc{\ec}{\end{center}}
\newc{\dg}{\dagger}
\newc{\prh}{\mbox{PR}_H}
\newc{\prq}{\mbox{PR}_q}
\newcommand{\tr}{\mathrm{Tr}}
\newc{\pd}{\partial}
\newc{\qv}{\vec{q}}
\newc{\pv}{\vec{p}}
\newc{\dqv}{\delta\vec{q}}
\newc{\dpv}{\delta\vec{p}}
\newc{\mbq}{\mathbf{q}}
\newc{\mbqp}{\mathbf{q'}}
\newc{\mbpp}{\mathbf{p'}}
\newc{\mbp}{\mathbf{p}}
\newc{\mbn}{\mathbf{\nabla}}
\newc{\dmbq}{\delta \mbq}
\newc{\dmbp}{\delta \mbp}
\newc{\T}{\mathsf{T}}
\newc{\J}{\mathsf{J}}
\newc{\sfL}{\mathsf{L}}
\newc{\C}{\mathsf{C}}
\newc{\B}{\mathsf{M}}
\newc{\V}{\mathsf{V}}

\begin{document}

%\newpage  
%\vspace{0.3cm}

%\begin{center}

\title{Diagonal unitary entangling gates and contradiagonal quantum states}
%:\\ Introducing the unimodular ensemble of random matrices}

\author{Arul Lakshminarayan}
\email[]{arul@physics.iitm.ac.in}
%\homepage[]{Your web page}
%\thanks{}
%\altaffiliation{}
\affiliation{Department of Physics, Indian Institute of Technology Madras, Chennai 600036, India}
\author{Zbigniew Pucha\l{}a}
\email{z.puchala@iitis.pl}
%{\large Karol \.Zyczkowski}  \footnote{Email address: karol@tatry.if.uj.edu.pl} \\
\affiliation{Institute of Theoretical and Applied Informatics, Polish Academy of Sciences, Ba{\l}tycka 5, 44-100 Gliwice, Poland}
\affiliation{Institute of Physics, Jagiellonian University, ul.\ Reymonta 4, 30-059 Krak\'ow, Poland}
%\altaffiliation{Center for Theoretical Physics, Polish Academy of Sciences, Warsaw, Poland}
\author{Karol \.Z{}yczkowski}
\email{karol@tatry.if.uj.edu.pl}
\affiliation{Institute of Physics, Jagiellonian University, ul.\ Reymonta 4, 30-059 Krak\'ow, Poland}
\affiliation{Center for Theoretical Physics, Polish Academy of Sciences, Warsaw, Poland}

\date{July 4, 2014}

\begin{abstract}
Nonlocal properties of an ensemble of diagonal random
unitary matrices of order $N^2$ are investigated.
The average Schmidt strength of such a bipartite diagonal quantum gate
is shown to scale as $\log N$, in contrast to the $\log N^2$ behavior
characteristic to random unitary gates. Entangling power of a diagonal gate $U$
is related to the von Neumann entropy of an
auxiliary quantum state $\rho=AA^{\dagger}/N^2$,
where the square matrix $A$ is obtained by reshaping the vector 
of diagonal elements of $U$ of length $N^2$
into a square matrix of order $N$.
This fact provides a motivation to study
the ensemble of non-hermitian unimodular matrices $A$,
with all entries of the same modulus and random phases
and the ensemble of quantum states $\rho$, 
such that all their diagonal entries are equal to $1/N$.  
Such a state is contradiagonal with respect to the computational basis, 
in sense that among all unitary equivalent states it maximizes the
entropy copied to the environment due to the coarse graining process.  
The first four moments of the squared singular values of the unimodular ensemble are derived,
based on which we conjecture a connection to a recently studied combinatorial object called the ``Borel triangle". 
This allows us to find exactly the mean von Neumann entropy for random phase density matrices
and the average entanglement for the corresponding ensemble of bipartite pure states.
\end{abstract}

\maketitle

\section{Introduction}
 \vspace{5mm}

Entanglement has been at the focus of recent researches in quantum information 
-- for a review see \cite{HHHHRMP09}, 
as it enables a range of uniquely quantum tasks such as teleportation, quite apart from 
being a singular nonclassical phenomenon and therefore of fundamental interest. It is well appreciated now that 
many particle pure states are typically highly entangled 
\cite{Lu78,Pa93,ZS01,Hayden06}, and share entanglement in a manner that is
 almost wholly of a multipartite nature. 
Here typicality refers to ensembles of pure states selected according to 
 the uniform (Haar) measure. If two distinct subsystems of a pure state are such that together they make up the entire
 system in a typical pure state, then the two subsystems will be largely entangled.
 In early works, Page and others \cite{Pa93,Sen96} had
 found the average entanglement, which in this case is simply the mean von Neumann 
entropy of the reduced state, and showed that it is nearly the maximum possible.

 More recent studies have explored the distribution of entanglement in such complete
 bipartite partitions of random states \cite{Nadal11}. If the two subsystems do not
 comprise the entire systems, for example two particles in a three particle state,
 the average entanglement depends on the dimensionality of the subsystems. 
Roughly speaking if the complementary space of the subsystems (say $A$ and $B$) is smaller, the density matrices 
$\rho_A$ and $\rho_B$ will be typically negative under partial transpose and therefore $A$ and $B$ are entangled \cite{Uday03}.

Ways to generate entangled states from initially unentangled ones via unitary operators is of natural interest, and in the context of quantum computation implies the construction of appropriate gates. Investigations of entangling power of a unitary quantum gate were initiated by Zanardi and co-workers \cite{ZZF00,Za01}, while some measures of non--locality were
analyzed in  \cite{HVC02,ZVWS02,WSB03,NDDGMOBHH02,Sc04,BS09}.
A typical quantum gate acting on a composed system consisting of
two $N$--level systems can be represented by a random unitary matrix of order $N^2$. Nonlocal properties of such random gates were investigated in \cite{MKZ13}.

In this work we shall analyze a simpler ensemble
of diagonal unitary random matrices and will
characterize nonlocality of the corresponding quantum gates.
It is also naturally related to an ensemble of pure bipartite  states in $N^2$ 
dimensional space whose components in some fixed basis are of the form $e^{i \phi_j}/N^2$, and $\phi_i$ are 
uniformly distributed random numbers.  Such an ensemble has been recently  studied as phase-random states \cite{Nakata2012}, 
and in fact connections to diagonal quantum circuits has been pointed out \cite{Nakata2014}. 
Part of the motivation for the study of diagonal quantum gates is that many Hamiltonians 
 have the structure that the basis in which the interaction is diagonal can be chosen to be unentangled. 
Time evolution is then governed by unitary operators whose entangling parts are diagonal.
 Recently studies of measurement-based quantum computation has also used diagonal unitary gates and shown that it can 
still remain superior to classical computation \cite{Bremner2011,Hoban2014}.

Also explicitly, unitary operators such as $(U_A\otimes U_B) U_{AB}$ occur in the study of coupled systems 
including kicked quantum tops -- see the book of Haake \cite{Haake}. 
The coupling could be of the form $J^z_AJ^z_B$, where $J_z^{A,B}$ are spin operators. 
Thus the nonlocal part of the evolution is diagonal again. It is found numerically that the
 eigenstates of such operators as well as the time evolution engendered by repeated applications 
of such operators can create large entanglement well approximated  by that of random states \cite{JNBL02}. 
However such operators have much fewer number of possible independent elements than Haar
 distributed unitaries on ${\cal H}^N \otimes {\cal H}^N$.
 Thus it is of interest to study  the origin of such large entanglement.

Furthermore, we are going to study the related ensemble 
of ``unimodular random matrices", comprising 
complex matrices whose all entries have the same modulus and
randomly chosen phases. Such matrices arise 
from reshaping a pure phase random state as defined in \cite{Nakata2012}.
Note that the usage of the term  'unimodular' concerns all the entries of a matrix,
so such a matrix is {\it not} unimodular in the sense of being integer matrices with determinants $\pm1$.

Although the unimodular ensemble differs from the Ginibre ensemble of
complex, non-hermitian matrices, with independent, normally distributed elements, it displays the same asymptotic behavior of the level density, 
which covers uniformly the unit disk.
% The spectrum of these non-hermitian matrices
%asymptotically covers a disk in the complex plane in agreement with the circular law. 
On the other hand the squared singular values of unimodular matrices
coincide with eigenvalues of certain specific quantum states of size $N$, 
the diagonal entries  which are equal to $1/N$. As the notion of 
an ``antidiagonal matrix" has entirely different meaning,
the density matrices with all diagonal elements equal 
will be called {\sl contradiagonal}. 
Observe that reduced density matrices of random phase pure states are thus contradiagonal.
We investigate properties of such an ensemble of quantum states
and discuss the contra-diagonalization procedure, which brings any hermitian matrix
to such a basis, that all their diagonal elements do coincide.

One may expect that random contradiagonal states correspond to the
large entanglement of the pure bipartite states 
and we show that indeed these states have larger von Neumann entropy
than those sampled according to the Ginibre ensemble. 
The unimodular  ensemble, while having no obvious invariance properties,   seems to also have remarkable underlying mathematical structure. For example, the average of the moments of the matrices in the 
ensemble are connected to polynomials with combinatorial interpretations. 
We evaluate the first four moments and use this to conjecture an exact expression for {\it all} of 
them. We numerically show that this is more than likely to be correct. 
Analytical continuation of the moments to non-integer powers allows us to evaluate the 
average von Neumann entropy for this ensemble which appears to be {\it exact} for all dimensions.

This paper is organized as follows. In Section II the ensemble of diagonal 
unitary gates is introduced, while in Section III the related ensemble of unimodular matrices is
analyzed.  Low order moments are  calculated analytically,
and a natural conjecture is made for exact expressions for all moments. 
Numerical evidence is presented for this. 
Nonlocality and entangling powers of random unitary gates is studied in Sec. IV. 
An exact expression based on a continuation of moments is presented for the average Schmidt 
strength of diagonal unitaries, or equivalently of the von Neumann entropy of the random phase states.
Procedure of contra-diagonalization of a hermitian matrix, which makes all diagonal
elements equal, is introduced in Sec. V. In this section we study in particular
the cognate ensemble of random contradiagonal states and show their particular
properties concerning the transfer of quantum information. The paper is concluded in Sec VI, and an 
Appendix reviewing the operator Schmidt decomposition and entangling entropy. 

\bigskip 

\section{Diagonal bipartite quantum gates}

Consider a diagonal unitary matrix $U$ of order  $N^2$.
Each entry is assumed to be random, so that 
$U_{\mu \nu}=\delta_{\mu \nu}\exp(i \phi_{\nu})$, 
where $\phi_{\nu}$ are independent random phases 
distributed uniformly in $[0,2\pi)$. Such a matrix
represents a diagonal unitary gate acting on a 
bipartite quantum system, a state in ${\cal H}^N \otimes {\cal H}^N$.

Any matrix $U$ acting on the composed Hilbert space ${\cal H}_N \otimes {\cal H}_N$,
 can be represented in its operator Schmidt form,
\begin{equation}
  U = \sum_{k=1}^{K} \sqrt{\Lambda_{k}} \; B_k^{\prime} 
  \otimes  B_k^{\prime\prime} ,
\label{VSchmidt}
\end{equation}
where the Schmidt rank $K \le N^2$. Note that the matrices
$B_k^{\prime}$  and $B_k^{\prime\prime}$ of order $N$ 
in general are non unitary.

It can be shown \cite{ZB04} 
that the Schmidt coefficients $\Lambda_k$, $k=1,\dots, K$,
can be obtained as squared singular values of the reshuffled matrix $U^R$.
This fact is briefly recalled in Appendix A,
where the notation is explained.
A generic diagonal matrix of size four,
after reshuffling forms a non-hermitian matrix
of rank two, 
\begin{equation}
U = \left[\begin{array}{cccc}
                  U_{11}  & 0 & 0 & 0 \\
                  0  & U_{22} & 0 & 0 \\
                  0  & 0 & U_{33} & 0 \\
                  0  & 0 & 0 & U_{44} \\
\end{array}\right], \quad
U^R = \left[\begin{array}{cccc}
                  U_{11}  & 0 & 0 & U_{22} \\
                  0  & 0 & 0 & 0 \\
                  0  & 0 & 0 & 0 \\
                  U_{33}  & 0 & 0 & U_{44} \\
\end{array}\right] . \quad
\label{u4} \
\end{equation}
For a diagonal matrix $U$ of order $N^2$
the reshuffled matrix $U^R$ contains $N(N-1)$ columns and rows
with all entries equal to zero. Hence the non-zero singular values of $U^R$
are equal to the singular values of a square matrix $A$ of size $N$,
obtained by reshaping the diagonal of the unitary gate,
  $A_{jk}=\exp(i \phi_{\nu})$, where $\nu=(j-1)N+k$.
As all entries of $A$ are unimodular and have a random phase, 
this construction defines an ensemble of random
unimodular matrices.

\section{Random unimodular matrices}

Consider a complex square matrix $A$ of size $N$ from the
{\sl unimodular ensemble}, so that a) all entries have the same modulus, $|A_{jk}|=1$,
and b) the phases are drawn independently from a uniform distribution,
\begin{equation}
 A_{jk} \ = \ \exp ( i \phi) ,  \quad P(\phi)=
\frac{1}{2\pi}, {\rm \quad for \quad}
\phi \in [0,2\pi).
\label{unimod}
\end{equation}
Such a random matrix $A$ could be called a {\sl pre--Hadamard}
as all entries have the same modulus, so choosing
an appropriate set of phases it may become unitary,
and thus belong to the class of complex Hadamard matrices \cite{TZ06}.
In our model all phases are random and non-correlated, so a
typical matrix from this ensemble exhibits effects of 
strong non-unitarity.
The ensemble of unimodular matrices  $A$ with all independent, 
well--behaved, identically distributed entries 
is of the Wigner type. Thus this non-hermitian ensemble
or random matrices satisfies asymptotically the circular
law of Girko \cite{TV11}.

\begin{figure}[htbp]
        \centerline{ \hbox{
        \epsfig{figure=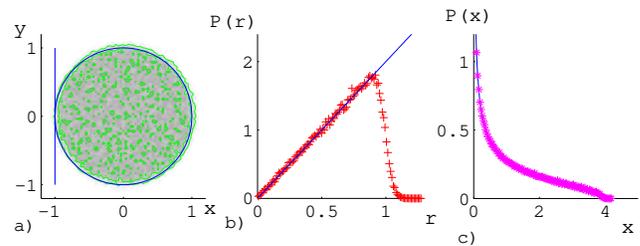,width=8.40cm} }}
     \caption{Properties of random unimodular matrices 
       of order $N=100$
     are close to those of the complex  Ginibre ensemble:
    a) the spectrum of $A/\sqrt{N}$ in the 
    complex plane satisfies the
     circular law of Girko; b) the radial density $P(r)$
     of the complex eigenvalues grows linearly 
    in the center of the unit disk;
    and c) the distribution of squared singular values $P(x)$
    is described by the Marchenko--Pastur distribution.}
     \label{fig_100}
\end{figure}

As shown in Fig.~\ref{fig_100} already for $N=100$ 
the spectral density for the unimodular ensemble
is close to uniform in the unit disk.
Furthermore, the distribution of rescaled 
squared  singular values, $x={\rm eig} (AA^{\dagger})/N$,
is asymptotically described by the Marchenko--Pastur  (MP) distribution
$P_{MP}(x)= \frac{1}{\pi}\sqrt{1/x-1/4}$ for $x \in [0,4]$,
characteristic to the Ginibre ensemble.
The moments of this distribution are given by the Catalan
numbers, while its entropy reads
 $-\int_0^4 x \log x P_{MP}(x)dx=-1/2$.
As $x=N \lambda$, where $\lambda$ denotes the eigenvalue of 
a normalized density matrix 
\begin{equation}
\label{AAdag}
\rho=AA^{\dagger}/N^2
\end{equation}
satisfying $\tr \rho=1$,
the average entropy of spectrum of $\rho$
behaves asymptotically as $\log N-1/2$.
Note that this behavior is characteristic
to random quantum states distributed uniformly
with respect to the Hilbert--Schmidt measure \cite{SZ04}
in the entire set of quantum states of a given dimension.

\begin{figure*}[htbp]
        \centerline{ \hbox{
              \epsfig{figure=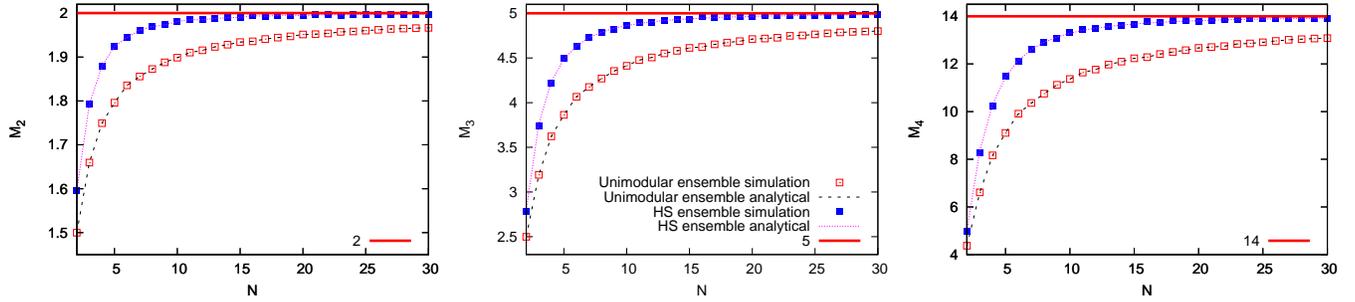,width=\linewidth} 
}}
        \caption{Moments of the distribution of squared singular
    values $P(x)$ for unimodular ensemble and the Hilbert--Schmidt
     measure as a function of the matrix size $N$: from left to right are shown the
  second moments $M_2$, third moments $M_3$, and fourth moments $M_4$ respectively.
   Horizontal lines at $M_2=C_2=2$ and $M_3=C_3=5$ and $M_4=C_4=14$
   denote the Catalan numbers,  which give the corresponding moments
 of the MP distribution and determine the the asymptotic behavior
 of both moments. Solid lines represent predictions (\ref{momHS})
for the Hilbert-Schmidt measure, while the dashed lines the predictions for the unimodular ensemble.
}
     \label{fig_moment}
\end{figure*}

Although for large $N$ statistical properties
of the unimodular ensemble coincide with those of 
complex Ginibre ensemble, deviations are visible for small matrix size. 
To visualize these effects we studied the moments of the 
distribution of squared singular values
$M_m =\int x^m P(x)dx$. Fig.~\ref{fig_moment}
shows a comparison of the moments $M_2$, $M_3$ and $M_4$ 
for random matrices of the unimodular ensemble
and the Hilbert--Schmidt ensemble of order $N$.
In the later case analytical predictions for the moments
are known as the traces of the random states
$\rho_N$ of size $N$ distributed according to the 
Hilbert--Schmidt measure read \cite{Lu78,SZ04},
\begin{equation}
\br {\tr} \rho_N^2\kt_{HS} = \frac{2N}{N^2+1} , \quad
\br {\tr} \rho_N^3 \kt_{HS}= \frac {5N^2+1}{(N^2+1)(N^2+2)} ,
\label{momHS}
\end{equation}
and due to rescaling of the variable $x$
one has  $M_m=N^{m-1} {\tr} \rho^m$.
Numerical data, such as that presented in Fig.~(\ref{fig_moment}), show that for a given $N$
the moments for the unimodular ensemble are smaller,
so the corresponding distribution are narrower,
even though for large $N$ both distributions tend to the
limiting Marchenko--Pastur distribution.

Note that the averages moments for the distribution of
squared singular values for random Ginibre matrices of a given size $N$,
derived recently in \cite{AKW13}, coincide with the predictions  (\ref{momHS})
for the HS ensemble only in the asymptotic case $N\to \infty$. In the former ensemble
the constraint concerns the average trace $\langle {\tr} GG^{\dagger}\rangle$, 
while in the latter each random matrix has a fixed trace, $\tr \rho=1$,
so that this difference  asymptotically vanishes.

\begin{figure*}[htbp]
        \centerline{ \hbox{
              \epsfig{figure=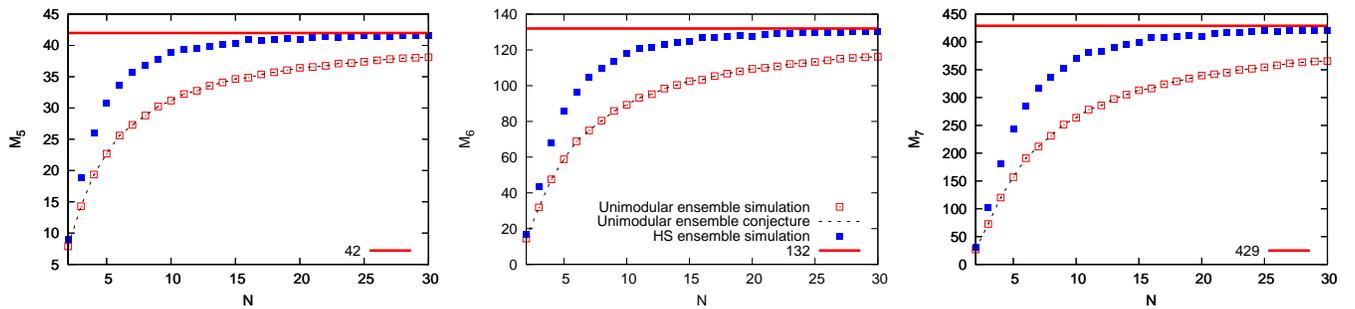,width=\linewidth} 
}}
        \caption{The fifth, sixth and seventh moments, $M_5$, $M_6$, and $M_7$ of the
 distribution of squared singular values $P(x)$ for the unimodular ensemble as a
 function of the matrix size $N$. the points are from numerical simulation based on a
 million realizations while the curves are from the formula in Eq.~(\ref{Eq:tr5}).
 The horizontal lines are at $C_5=42$, $C_6=132$ and $C_7=429$, further Catalan numbers
 and the asymptotic value of the scaled moments.
}
     \label{fig_tr5}
\end{figure*}

\subsection{Lower order moments and a conjecture for all orders}

The lower order moments of the unimodular ensemble can be exactly evaluated and compared to the above case. 
In fact we will calculate exactly moments till the fourth and conjecture an exact formula for any moment.
The density matrix elements are
\beq
\rho_{\alpha_1 \alpha_2}=\df{1}{N^2}\sum_{l_1=1}^{N}\exp[i(\phi_{\alpha_1 l_1}-\phi_{\alpha_2 l_1})].
\label{momUni}
\eeq

\subsubsection{ The second moment $\br \tr \rho_N^2 \kt_{UE} $}
The second moment while being the simplest, also serves as a measure of purity of a given quantum mixed state. We have 
\beq
\tr \rho^2= \df{1}{N^4} \sum_{\alpha_1 \alpha_2}\sum_{l_1 l_2}\exp[ i(\phi_{\alpha_1 l_1}-\phi_{\alpha_1 l_2}+\phi_{\alpha_2 l_2}-\phi_{\alpha_2 l_1})].
\label{trrho2A}
\eeq
On averaging over the uniform phases the only terms that would survive are those cases for which the phase vanishes. This happens if $\alpha_1=\alpha_2$ for arbitrary $l_1$ and $l_2$. Thus this case contributes 
\beq
\df{1}{N^4}\sum_{\alpha_1}\sum_{l_1l_2} \, 1=\df{N^3}{N^4}=\df{1}{N}.
\eeq
Indeed this is the ``diagonal" contribution. The phase also vanishes if $\alpha_1 \neq \alpha_2$, but $l_1=l_2$. This ``off-diagonal" contribution is 
\beq
\df{1}{N^4}\sum_{\alpha_1\neq \alpha_2}\sum_{l_1}\, 1= \df{1}{N^4}N(N-1)N=\df{N-1}{N^2}.
\eeq
As these exhaust the exclusive possibilities, the average second moment for the
level density of related to the unimodular ensemble reads
\beq
\br \tr \rho_N^2 \kt_{UE} = \df{2N-1}{N^2}.
\label{trrho2B}
\eeq
Equivalently $M_2=(2N-1)/N$, which indeed tends to $2$ for large $N$. Also note that $\br \tr \rho^2 \kt_{HS}-\br \tr \rho^2 \kt_{UE}=(N-1)^2/[N^2 (N^2+1)]>0$, 
indicating that the density matrices constructed from the unimodular ensemble are on average more mixed than those from the Ginibre ensemble.

\subsubsection{ The third moment $\br \tr \rho_N^3 \kt_{UE} $}

The third moment is obtained by considering all those cases when
\[ \phi_{\alpha_1 l_1}-\phi_{\alpha_1 l_2}+\phi_{\alpha_2l_2}-\phi_{\alpha_2 l_3}+\phi_{\alpha_3 l_3} -\phi_{\alpha_3 l_1} \]
vanishes for arbitrary sets of phases, where $\alpha_i$ and $l_i$ take values in $1, \cdots , N$. The indexes are ordered in such a way that a bijection to a standard counting problem becomes possible. Starting from the first pair, the sign is reversed while the second index is new in the second pair. The third pair is obtained by again reversing the sign of the second but now the first index becomes new, and so on. Finally when there are $6$ pairs, the last index is the same as that of the first pair. It is clear that this generalizes to the moment of order $k$, where there are $2k$ such pairs.

At this level a bijection to several standard problems in counting that involve the Catalan numbers $C_k$ \cite{Koshy} is possible. For example that of matching 3 pairs of parentheses: $()()(),(())(),()(()),((())),(()())$ is relevant to the third moment,
each parenthesis represents the pair of indexes at the corresponding place.
The matched pair of parentheses imply that the pair of indexes are equal. 
The first possibility and its translation in terms of indexes is: $()()()$: $\alpha_1 l_1=\alpha_1 l_2$,
$\alpha_2 l_2=\alpha_2 l_3$, $\alpha_3 l_3=\alpha_3 l_1$, or $l_1=l_2=l_3$. The second $(())():$ $\alpha_1 l_1=\alpha_2 l_3$, $\alpha_1 l_2=\alpha_2 l_2$, $\alpha_3 l_3=\alpha_3 l_1$, or $\alpha_1=\alpha_2$, $l_1=l_3$. Similarly $()(())$: $\alpha_2=\alpha_3$, $l_1=l_2$, $((()))$: $\alpha_1=\alpha_3$, $l_2=l_3$, $(()())$: $\alpha_1=\alpha_2=\alpha_3$. The unconstrained indexes can take arbitrary values between $1$ and $N$. Another bijection is between the indexes and points that are joined by noncrossing semicircles. Thus the contributions from $()()()$ and $(()())$ are respectively
\[ \sum_{l_1}\sum_{\alpha_1,\alpha_2,\alpha_3} 1= N^4,\; \sum_{\alpha_1}\sideset{}{'}\sum_{l_1,l_2,l_3} 1= N^4-N^2.\]
The last sum is restricted in the sense that it does not include the case $l_1=l_2=l_3$ that is already included in the first count. The other three cases contribute equally 
\[ \sum_{\alpha_1\ne \alpha_3} \sum_{l_1\ne l_2}1=N^2(1-N)^2,\]
where the distinct indexes are always unequal; if they are equal it will reduce to a term considered in the first two cases.
Thus putting them all together we get 
\beq
\br \tr \rho^3 \kt_{UE} = \frac{1}{N^4}(5N^2-6N+2),
\eeq
and $M_3=N^2 \br \tr \;\rho_N^3 \kt =(5-6/N+2/N^2)$.

The $k$--{th} moment $\br \tr \rho^k\kt$ is of the form $P_k(N)/N^{2(k-1)}$,
where $P_k(N)$ is a degree $k-1$ polynomial in $N$ whose leading term's coefficient is $C_k=\frac{1}{k+1}\binom{2k}{k}$, the $k$--{th} Catalan number. Thus it follows that the moments $M_k=N^{k-1} \br \tr \rho^k\kt$ tend to $C_k$ and hence the asymptotic density is described by the universal Marchenko-Pastur distribution. 
Writing explicit expressions for the moments $M_k$ for the
unimodular ensemble we are in position to quantify the deviations from the
asymptotic universal MP distribution.
%That the leading term in $M_k$ is the Catalan number follows for $k>3$ as well is
% derived below. Corrections to the will become clear as explicit expressions 
%for the moments are written down.

\subsubsection{ The fourth moment $\br \tr \rho_N^4 \kt_{UE} $}
We will now evaluate explicitly the fourth moment that presents some challenges and then {\it conjecture} an {\it exact} expression for $P_{k}(N)$. There are 14 different parenthesizations for the $k=4$ case with $8$ pairs of indexes involved. There are 3 ``contractions" in each case. For instance the contractions corresponding to $()(())()$ are $l_1=l_2$, $\alpha_2=\alpha_3$, $l_4=l_1$. Thus there are $8-3=5$ sums that are unconstrained from each of the $14$ parenthesizations and hence the leading term in $N^8 \br \tr \rho_N^4\kt$ is $14 N^5$. Generalizing, to the $k$--th moment, the number of contractions is $k-1$. This is seen by writing the alternating indexes as actual  products and sums while respecting their distinctness and requiring $\sum_{i=1}^k \alpha_{i}(l_i-l_{i+1})=0$ with $l_{k+1}=l_1$. Setting say $\alpha_{n}l_n =\alpha_{m}l_{m+1}$
for some $n$ and $m$ reduces the number of terms by $1$. Due to periodic boundary conditions, the number of contractions that will set the whole sum to 0 is $k-1$. Thus it follows that in general the leading term in $N^{2k} \br \tr \rho^k\kt$ is $C_k N^{2k-(k-1)}=C_k N^{k+1}$. 

Overcounting however lowers the value of  $N^8 \br \tr \rho_N^4\kt$ from $14 N^5$, and in general from $C_k N^{k+1}$. Sticking to the $k=4$ case, the combined contributions from $()()()()$ and $(()()())$ which either contract all $l_i$ or all  $\alpha_i$ respectively is $2N^5-N^2$, $N^2$ being the double count of all $\alpha_i$ being the same and all $l_i$ being the same. The other $12$ contributions involve sums such as 
\[ \sum_{\alpha_1=\alpha_2}\sum_{\alpha_3=\alpha_4 \ne \alpha_1}\sum_{l_1=l_3} \sideset{}{'}\sum_{l_2,l_4}1= (N^2-1)(N-1)N^2,\] where the restricted sum eliminates the cases when $l_2=l_4=l_1=l_3$, which has already been considered. While this case corresponds to the parenthesization $(())(())$,
the other $11$ have sums over 5 indexes with similar restrictions. 

Thus these contribute $12N^2(N-1)(N^2-1)$, however there still remains some overcounting to be accounted for. For example the above case includes instances when $l_2=l_1=l_3$ but $\ne l_4$ which is also possible when the contracted indexes are $\alpha_2=\alpha_4$, and $l_1=l_2=l_3$ and corresponds to the paranthesization $()()(())$. Exhaustive enumeration of these terms that originate from contracting $4$ indexes reveals that there are $16$ such terms each of which gives
\[ \sum_{\alpha_1=\alpha_2}\sum_{\alpha_3=\alpha_4\ne \alpha_1} \sum_{l_1=l_2=l_3}\sum_{l_4\ne l_1} 1= N^2(N-1)^2.\]
There is no further overcounting as the cases with $>4$ contractions have already been properly included. Finally in total the contribution is $2N^5-N^2+12(N^2-1)(N-1)N^2 - 16 N^2(N-1)^2=14N^5-28 N^4 +20N^3 -5N^2$ and we get 
\beq
\br \tr \; \rho_N^4\kt_{UE}=\frac{1}{N^6}(14N^3-28 N^2+20N-5).
\eeq

\subsubsection{ A conjecture for all moments $\br \tr \rho_N^n \kt_{UE} $}

The complexity of the counting problem is naturally increasing. However having found the polynomials $P_2(N)=2N-1$, $P_3(N)=5N^2-6N+2$ and $P_4(N)=14N^3-28N^2+20N^3-5$ exactly, the following are evident: they have alternating signs, satisfy $P_k(1)=1$, and the constants (coefficients of $N^0$) have the absolute value $1$, $2$ and $5$ which are themselves Catalan numbers. That $P_k(1)=1$ follows simply as for $N=1$, there is a only a single pure phase $e^{i \phi}$ and the ``density matrix" is simply $e^{i \phi} e^{-i \phi}=1$.

Based on the triangle of coefficients $\{1,\{2,1\},\{5,6,2\},\{14,28,20,5\}\}$ a search in the On-line Encyclopedia of Integer 
Sequences \cite{Sloane} (OEIS) returns two sequences A062991 and A234950.
In the first of which is the signed version (that we encounter), it arises as generalizations of Pascal's triangles via 
Riordan arrays \cite{Barry2013}. In the second the unsigned version, which the authors call {\sl Borel's triangle} \cite{Francisco2013} 
and make connections to new counting problems in commutative algebra and discrete geometry.
 Going beyond what we have above, the first seven rows of the unsigned triangle reads:
\[
\begin{array}{lllllll}
1&&&&&&\\
2&1&&&&&\\
5&6&2&&&&\\
14&28&20&5&&&\\
42&120&135&70&14&&\\
132& 495& 770& 616& 252& 42&\\
429&2002& 4004& 4368& 2730& 924& 132\\
\ldots&&&&&& \end{array}
\]
The left and rightmost entries are Catalan numbers. The entry $f_{n,k}$ ($n\ge 0,\, k \ge 0)$ of the Borel
triangle is 
\beq
f_{n,k}=\sum_{s=0}^{n} \binom{s}{k} C_{n,s}
\eeq
where $C_{n,s}$ is {\sl Catalan's triangle}:
\[ 
\begin{array}{lllll}
1&&&&\\
1&1&&&\\
1&2&2&&\\
1&3&5&5&\\
1&4&9&14&14\\
\ldots&&&& \end{array}
\]
which satisfies the recursion $C_{n,k}=C_{n-1,k}+C_{n,k-1}$, that is the entries are sums of the one to the left and the one above. The first column of all $1$ is the ``boundary condition" $C_{n,0}=1$. Explicit formula for $C_{n,k}$ and  $f_{n,k}$ are available \cite{Barry2013}:
\[
C_{n,k}=\dfrac{(n+k)!(n-k+1)}{k!(n+1)!},\] \[  f_{n,k}=\frac{1}{n+1}\binom{2n+2}{n-k}\binom{n+k}{k}.
\]
It is then natural to {\it conjecture} that for $n\ge 1$
the average moments of the level density of the random matrices of size $N$
are 
\beq
\label{momentUE}
\br \tr \rho_N^n \kt_{UE}= \frac{1}{N^{2(n-1)}}\sum_{k=0}^{n-1} (-1)^{k} f_{n-1,k}\,N^{n-k-1}.
\eeq
The smallest unproven case is $n=5$, which can be simply read off from Borel's triangle:
\beq
\br \tr \rho^5 \kt_{UE} = \frac{1}{N^8}( 42N^4-120N^3+135 N^2-70N+14).
\label{Eq:tr5}
\eeq
Fig.~\ref{fig_tr5} displays the numerical 
calculations of moment $M_5=N^4 \br \tr \rho^5 \kt $, as well as $M_6$ and $M_7$ from a million random realizations of the ensemble and shows how well this conjecture fares. There seems to be no room for doubting the correctness of the conjecture.

A calculation of the first few cumulants from the moments $M_n$ of the scaled variables $N \lambda$ results in (apart from $\kappa_1=1$)
\beq
\begin{split}
 \kappa_2=\frac{1}{N}(N-1), \\ \kappa_3=\frac{1}{N^2}(N-1)(N-2), \\\kappa_4=-\frac{1}{N^3}(N-1)(4N-5), \\ \kappa_5=-\frac{1}{N^4}(N-1)(N-2)(4N^2+2N-7).
\end{split}
\eeq
As $N\rarrow \infty$ these tend to $\{1,1,1,0,-4, \ldots\}$, the initial cumulants corresponding to the moments being the Catalan numbers $\{1,2,5,14,42, \cdots\}$. 

The moments of the unimodular ensemble themselves also seem to have combinatorial significance. For example the moments for $N=3$ are such that $3^{(n-1)}M_n(N=3)$ is the integer sequence $\{1,5,29,181, 1181, \cdots\}$. If we include an additional $1$ corresponding to $M_0$, this sequence is found as a column in the entry A183134 of  the OEIS.
Indeed the other columns of the square array of this entry are similarly the moments for different values of $N$, $N=1,2, \cdots$. 
This prompts the additional conjecture that the $N^{2(n-1)}\tr(\rho^n_{N}) $
is the same as the number of $N$-alphabet words of length $2n$ beginning with the first character of the alphabet by repeatedly inserting doublets into the initially empty word, as this is the counting problem that is stated in the OEIS entries (see also \cite{Kassel2013}). For example in the case of $n=2$ and $N=2$, the alphabet set is binary $\{ab,\}$. Doublets are repeated alphabets. Thus inserting two doublets (for $n=2$) one gets $aaaa$, $aabb$, $abba$ as the three possibilities that start with $a$. This coincides with $2N-1$ that we derived above.

Having explored the moments of the unimodular ensemble 
we now turn to one of our central motivations, finding the entangling power of diagonal unitaries.

\section{Entropy of the unimodular ensemble and nonlocality of random diagonal gates, }

By construction, the average entropy of squared singular values
of random unimodular matrices $A$ is equal to the average
entropy of entanglement for the corresponding unitary gates $U=A^R$.
As squared singular values of $A$ are asymptotically 
described by the Marchenko--Pastur distribution,
making use of the Page formula \cite{Pa93}
we infer that the mean entropy of entanglement  
(\ref{shannon}) of a random diagonal gate behaves as
\begin{equation}
\langle S(U) \rangle_{\rm diag}  \ = \ 
\langle S(\rho=AA^{\dag}/N^2) \rangle_{UE} \ \approx \ \log N-1/2.
\label{entropy1}
\end{equation}
This result forms approximately a half of the entropy of
 generic Haar random unitary matrices \cite{MKZ13}
\begin{equation}
\langle S(U) \rangle_{\rm Haar} 
\ \approx \ 2 \log N-1/2.
\label{entropy2}
\end{equation}

\begin{figure}[htbp]
        \centerline{ \hbox{
        \epsfig{figure=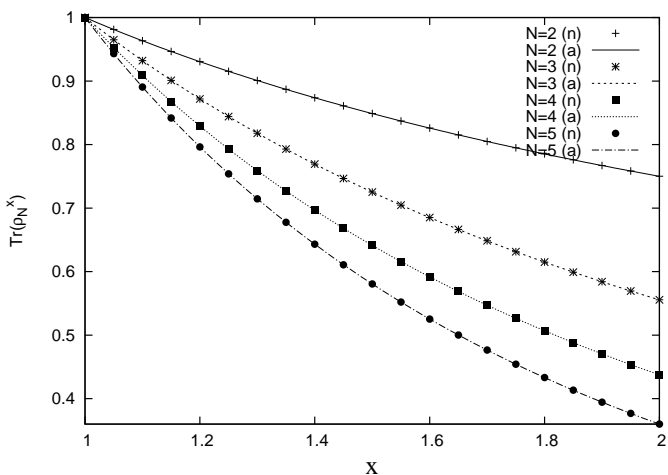,width=6.40cm,angle=-90} }}
     \caption{The moments for non-integer powers between $1$ and $2$ are plotted for the unimodular ensemble, and $N=2,3,4$. 
The points are obtained by numerical simulation using $10^6$ realizations from the ensemble, while the lines
 are those from the analytical expression in Eq.~(\ref{eq_momx}). }
     \label{fig_momx}
\end{figure}

Based on the discussion of moments in the previous section, and taking them to be exact allows us to find 
what appears to be an exact expression for the average entropy of entanglement,
 which can be interpreted as entanglement in a random ensemble of states $\br S(\rho)\kt_{UE}$ where $\rho$ 
is defined in Eq.~(\ref{AAdag}) or the average entanglement of diagonal unitary gates via Eq.~(\ref{shannon}).
 Using the view of state entanglement, from the last section, the expression in Eq.~(\ref{momentUE}) 
can be used to write the $n^{th}$  moment as 
\beq
\begin{split}
\br \tr \rho_N^n \kt _{UE}=\frac{1}{nN^{n-1}} \binom{2n}{n-1} \, _2F_1(n,1-n;2+n;1/N).
\end{split}
\eeq
Using this we can continue the moments to noninteger powers, so that we have $\sum_i \lambda_i^x$ for $x$ real. 
Observing that the average entropy $\br S(\rho)\kt_{UE}$ is the limit of $(1-\sum_i \lambda_i^x)/(x-1)$ as $x \rarrow 1^+$, we have that 
\beq
\br S(\rho)\kt_{UE} =-df(x)/dx \vert_{x=1} 
\eeq
where $f(x)=\br \tr \rho_N^x \kt=$
\beq
\frac{1}{N^{x-1}} \df{\Gamma(2x+1)}{\Gamma(x+1)\Gamma(x+2)} \, _2F_1(x,1-x;2+x;1/N).
\label{eq_momx}
\eeq
While it is not evident that such a continuation be exact, that it is indeed very likely to be 
so is illustrated in Fig.~(\ref{fig_momx}), where the moments are plotted for $1\le x \le 2$ for small values of $N$.
 This gives us confidence that the entropy found from such a procedure is also {\it exact}.

To evaluate $-f'(1)$ one needs to evaluate the derivatives of gamma functions and the hypergeometric function with respect to their parameters.
 Using that $\Gamma'(z)=\Gamma(z) \psi_0(z)$, where $\psi_0(z)$ is the digamma function we get that 
\beq
-\df{d}{dx} \df{1}{N^{x-1}}\df{\Gamma(2x+1)}{\Gamma(x+1) \Gamma(x+2)}|_{x=1}=\log N -\df{1}{2},
\label{eq_ent_der1}
\eeq
where the origin of $1/2$ is due to the fact that $\psi_0(3)-\psi_0(2)=1/2$. The derivative of the hypergeometric function can 
be evaluated using its definition as an infinite series. We have to compute 
\beq
\df{d}{dx} \sum_{m=0}^{\infty} \df{(x)_m (1-x)_m}{(x+2)_m} \df{1}{m! N^m},
\eeq
for which we need the derivatives of the Pochhammer symbols which are defined as $(a)_n =a (a+1) \cdots (a+n-1)$. Due to the fact that we need to evaluate the derivative at $x=1$ (and $(0)_m=0$ for $m>1$, while $(0)_0=1$), we only need that $d (1-x)_m/dx|_{x=1}=-(m-1)!$ which is easy to see from the definition of the symbol. Putting these together we get that 
\beq
\begin{split}
&-\df{d}{dx}\,  _2F_1(x,1-x,x+2;1/N)|_{x=1}=\\ &\sum_{m=1}^{\infty}\df{2}{m(m+1)(m+2)} \df{1}{N^m}= \\&\df{3}{2}-N-(N-1)^2 \log(1-\f{1}{N}).
\label{eq_ent_der2}
\end{split}
\eeq
The last equality is obtained as the  infinite sum can be evaluated by elementary means, integrating thrice the identity $1/(1-x)= 1+x+x^2+ \cdots$.
and thus finally the average entropy is the sum of the results in Eqs.~(\ref{eq_ent_der1}) and \ref{eq_ent_der2}: 
\beq
\br S(\rho)\kt_{UE}=\log N -(N-1)-(N-1)^2\log\left(\df{N-1}{N}\right).
\label{entropyUE}
\eeq

\begin{figure}[htbp]
        \centerline{ \hbox{
        \epsfig{figure=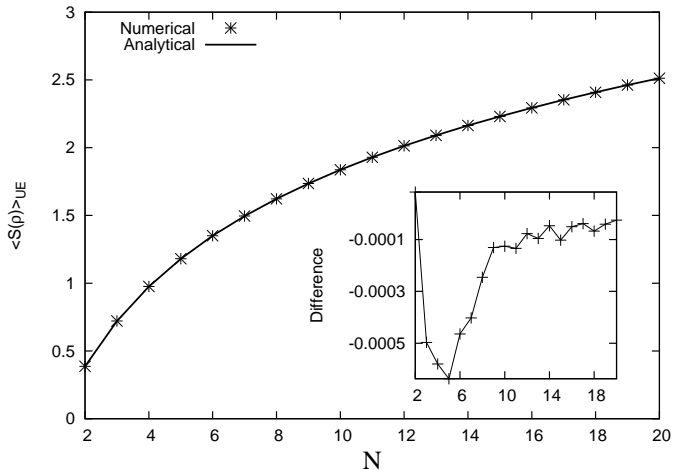,width=6.40cm,angle=-90} }}
     \caption{The average entropy or entanglement in the unimodular ensemble is plotted against various dimensionalities. 
The points are obtained by a numerical simulation using $10^6$ realizations, while the smooth line is from
 the analytical formula in Eq.~(\ref{entropyUE}). The essentially exact nature of the expression is illustrated in
 the inset where the difference between the numerical and analytical is shown to be consistent with statistical
 fluctuations from a million realizations. }
     \label{fig_ent}
\end{figure}

Numerical results presented in Fig.~(\ref{fig_ent}) provide further arguments
that the above expressions for the average entropy are exact for any dimension.
The differences between  the formula and numerical simulations are shown to be smaller than $1/\sqrt{N_S}$,
 where $N_S$ denotes the size of the numerical sample. 
For large $N$ it is easy to see that this approaches $\log N-1/2$ with the neglected terms being of order $1/N$,
 in agreement with what is expected from the Marchenko-Pastur distribution  and from the Page formula \cite{Pa93} for the HS ensemble.
 For instance, if  $N=2$, the exact density of the eigenvalues in $[0,1]$ reads $1/(\pi\sqrt{y(1-y)})$, as discussed in the next section.
Hence the average entropy is
\beq
\int_0^1 \df{-2y \log y}{\pi \sqrt{y(1-y)}}=\log 4 -1,
\eeq
which agrees with Eq.~(\ref{entropyUE}). We may compare this with the exact entropy from the HS ensemble,
 which is $\br S(\rho)\kt _{HS}=\sum_{k=N+1}^{N^2}1/k-(N-1)/(2N)$ \cite{Pa93}. 
For example for $N=2$ this gives $1/3$ which is smaller than that for the UE ensemble that is $\log 4-1 \approx 0.39$. 
Although the difference decreases with the matrix size, the relation
$\br S(\rho)\kt _{HS}<\br S(\rho)\kt _{UE}$ holds true, 
which indicates again the enhanced average entanglement in the unimodular ensemble.

Another measure of nonlocality of gates is the so-called {\sl entangling power}
 based on the ability of operators to create subsystem mixed states from originally
 unentangled pure bipartite states. If $|\psi_1\kt \otimes |\psi_2\kt $ is an
 unentangled state in ${\cal H}_N \otimes {\cal H}_N$, and $U$ is an unitary operator 
on this space, its entangling power as defined by Zanardi, Zalka and Faoro 
\cite{ZZF00} is
\beq
e_p(U)=\overline{E(U|\psi_1\kt \otimes |\psi_2 \kt)}^{\psi_!,\psi_2},
\eeq
the average being over all product states. Any entanglement measure can be used for $E$, the one that was used in \cite{ZZF00} 
being the simplest useful one, the linear entropy: $E(|\psi\kt)= 1-\tr_1 \mu^2,$ where  $\mu\equiv \tr _2 |\psi\kt \br \psi|$ 
is the reduced density matrix of the subsystem labeled by 1.

 The case of interest in the present work is one where the matrix $U$ is diagonal and hence the following is obtained:
\beq
\br \alpha | \br l |U|\psi_1 \kt |\psi_2\kt = e^{i \phi_{\alpha l}} \br \alpha|\psi_1 \kt \br l |\psi_2\kt, 
\eeq
where we have used $\br \alpha l |U |\beta m \kt = e^{i \phi_{\alpha l}} \delta_{\alpha \beta} \delta_{lm}$. The reduced density matrix is 
\beq
\mu_{\alpha_1 \alpha_2}=\sum_{l}e^{i (\phi_{\alpha_1 l}- \phi_{\alpha_2 l})} \br \alpha_1|\psi_1\kt \br l|\psi_2\kt \br \psi_1|\alpha_2\kt \br \psi_2|l \kt, 
\eeq
and therefore 
\beq
\begin{split}
\tr \mu^2 =\sum_{\alpha_1,\alpha_2,l_1,l_2}e^{i(\phi_{\alpha_1 l_1}-\phi_{\alpha_1 l_2}+\phi_{\alpha_2 l_2}-\phi_{\alpha_2 l_1})} \\
\times |\br \alpha_1|\psi_1\kt|^2 |\br l_1|\psi_2\kt|^2 |\br \alpha_2|\psi_1\kt|^2 |\br l_2 |\psi_2 \kt|^2.
\end{split}
\eeq
Averaging over the states $|\psi_{1,2}\kt$ can be done assuming that they are random vectors distributed uniformly according to the invariant Haar measure. Further assuming the general case of complex random states  \cite{Brody81}, the following are the average of the products of two intensity components:
\beq
\overline{|\br \alpha_1|\psi_1\kt |^2 |\br \alpha_2|\psi_1\kt|^2}=\dfrac{\delta_{\alpha_1 \alpha_2}+1}{N(N+1)}.
\eeq
Using this, it follows that 
\beq
\overline{\tr \mu^2}=\dfrac{N^2+2 N^3 +N^4\tr(\rho^2)}{N^2(N+1)^2}.
\eeq
The connection to $\tr \rho^2$ of the last section follows from Eq.~(\ref{trrho2A}). Thus the entangling power of diagonal unitaries $e_p(U)=1-\overline{\tr \mu^2}$ is directly related to the second moment of the squared singular values of the reshaped matrix. The average entangling power, now averaged over all phases in the diagonal unitary gates is 
\beq
\br e_p(U)\kt_{diag} = 1-\dfrac{N^2+2 N^3 +N^2(2N-1)}{N^2(N+1)^2}= \left(\frac{N-1}{N+1}\right)^2,
\eeq
where the result $\br \tr \rho^2\kt_{UE} = (2N-1)/N^2$ has been used from the previous section. This can be compared with the average entangling power of unitary gates \cite{ZZF00}: $\br e_p(U)\kt = (N-1)^2/(N^2+1)$, which is only marginally larger. The entropies of full unitaries were almost twice as large as the diagonal ones. At the level of the purity however one still sees the difference in that $\br \overline{\tr \mu^2}\kt=4N/(N-1)^2 \approx 4/N$ is double that of the reduced density matrix of typical random states in ${\cal H}_N \otimes {\cal H}_N$, which reads $2/N$ \cite{Lu78}.

\section{Contradiagonal hermitian matrices}

Any ensemble of random matrices, 
allows one to generate an ensemble of quantum states \cite{ZPNC11}.
Taking a random matrix $A$, 
one writes $\sigma=AA^{\dagger}/{\tr} AA^{\dagger}$
to get a random density matrix: a hermitian, positive operator
normalized by the trace condition ${\tr} \sigma=1$.
In the case of unimodular random matrices (\ref{unimod})
one has ${\tr} AA^{\dagger}=N^2$, 
hence $\sigma=AA^{\dagger}/N^2$.

Observe that by construction of the unimodular matrix $A$
the corresponding positive Wishart matrix $AA^{\dagger}$ 
has all diagonal elements {\it equal}. Thus the diagonal elements of the
corresponding random density matrix 
$\rho$ read $\rho_{ii}=1/N$, where $i=1,\dots,N$.
In other words, the state $\rho$ is represented
in such a particular basis $\{|1\rangle, |2\rangle, \dots, |N\rangle\}$
that the expectation values among each of the basis states are equal.
Thus the entropy of an orthogonal measurement in this basis
is maximal and equal to $\log N$. We show below that
such a basis is dual to the basis in which a given state is
diagonal, in sense that the norm of all the off-diagonal elements is
maximal. Thus any density matrix $\sigma=AA^{\dagger}/N^2$
constructed out of a random unimodular matrix $A$
has all diagonal elements equal
and therefore  will be called {\sl contradiagonal}.

\subsection{Procedure of contra--diagonalization of a matrix}

Let $H$ be a Hermitian matrix of order $N$ and let $G=VHV^{\dagger}$ be a
unitarily similar matrix, as $VV^{\dagger} = V^{\dagger}V = {\mathbb I}_N$. All
matrices from this orbit share the same spectrum  and posses the same trace,
${\tr} G = {\tr} H =:t$.
For any fixed $H$ we are going to analyze the sum of the squared moduli of
off-diagonal elements of $G$ and define a function $f(V)=\sum_{i \ne j}
|G_{ij}|^2$. Let us denote by $D$ any hermitian matrix $VHV^{\dagger}$ for which
the function $f$ becomes minimal, 
\begin{equation}
   D=U_{min}HU_{min}^{\dagger} \; : f(U_{min})= \min_{V \in U(N)} \sum_{i \ne j} |(VHV^{\dagger})_{ij}|^2 .
\label{minD}
\end{equation}
In  a similar way let $A$ represent a hermitian matrix $VHV^{\dagger}$
for which the function $f$ becomes maximal, 
\begin{equation}
   A=U_{max}HU_{max}^{\dagger} \; : f(U_{max})= \max_{V \in U(N)}  \sum_{i \ne j} |(VHV^{\dagger})_{ij}|^2  .
 \label{maxA}
\end{equation}
It is clear that the minimum $f(U_{min})=0$ is achieved for a matrix $U=U_{min}$ consisting of eigenvectors
of $H$, so $D$ is a diagonal matrix of eigenvalues of $H$ or any other matrix
similar with respect to permutations $D'=PDP^T$.
As the standard procedure to find $D=UHU^{\dagger}$ is called diagonalization,
the transformation of $H$ by $U_{max}$, leading to the maximum in (\ref{maxA}),
will be called {\sl contra--diagonalization}. 
For any given hermitian $H$ we shall show below how to find its contra--diagonalizing matrix
$U_{\rm max}$.
\medskip 

It is well known \cite{hj2} that any matrix can be unitarily transformed to a form, in
which all the diagonal entries are equal.  As the trace of a Hermitian matrix $H$ 
is unitarily invariant this constant reads $H_{jj}={\tr} H/N$ for $j=1,\dots, N$.

Let $F$ denote a complex Hadamard matrix \cite{TZ06} of order $N$, 
so that $F$ is unitary and  the moduli of all its entries are equal, 
$|F_{jk}|=1/\sqrt{N}$.
As a typical example let us mention the {\sl Fourier matrix} of size $N$ with entries
$(F_N)_{jk}=\exp(i  jk \pi/N)/\sqrt{N}$ for $j,k=0,\dots, N-1$. 
Two complex Hadamard matrices are  called equivalent, written $H_2 \sim H_1$, if they are equivalent
up to enphasing and permutations, $H_2=P_1E_2H_1 E_2P_2$. Here 
$E_1$ and $E_2$ denote diagonal unitary matrices
while $P_1$ and $P_2$ represent permutation matrices of order $N$.
For $N=2,3$ and $N=5$ all complex Hadamard matrices are 
equivalent to the Fourier matrices $F_2$, $F_3$ and  $F_5$ respectively,
 see  \cite{TZ06}.
\medskip

In order to compare spectra of hermitian matrices
it is convenient to use the notion of majorization.
Consider vectors of size $N$
ordered decreasingly, $x_1 \ge x_2 \ge \dots x_N$.
A vector  $y$ 
is said to {\sl majorize} \cite{MO79} vector $x$,
written $x \prec y$, if partial sums satisfy following
inequalities  $\sum_{i=1}^m {x}_i  \leq  \sum_{i=1}^m {y}_i$
for $m=1,\dots, N-1$ and additionally $\sum_{i=1}^N {x}_i  = \sum_{i=1}^N {y}_i$.
A function $f:\mathbb{R}^N \to R$ is said to be Schur convex when
$x \prec y$ implies $f(x) \leq  f(y)$.

Now we can formulate the following result.
\begin{proposition} 
\label{lemma:anti}
Let $H$ be a Hermitian matrix of order $N$ and let $V$ be unitary. Then 

a) the maximum in (\ref{maxA}) is obtained for $U_{\max}=F U_{min}$
where $U_{min}$ denotes the matrix of eigenvectors of $H$ and $F$ is a complex Hadamard matrix,

b) the matrix $  A=U_{max}HU_{max}^{\dagger}$ obtained in this way is contradiagonal, 
 $A_{jj}={\tr}H/N$ for $j=1,\dots, N$,
 
c) the maximum reads $f_{max}(V)={\tr} H^2 - ({\tr} H)^2/N $.
\end{proposition}

\proof[Proof of Proposition~\ref{lemma:anti}]

First to show item b) we consider first  diagonal matrix $H=D$
and take an arbitrary complex Hadamard matrix $F$ and find that 
$A = F D F^{\dagger}$  is contradiagonal, as all its diagonal elements are equal, 
% $A_{ii} = \frac{t}{N}$,
\begin{equation}
\begin{split}
(F D F^{\dagger})_{ii} = \sum_{k l} F_{ik} D_{kl} (F^{\dagger})_{l i} =
\sum_{k} F_{ik} D_{k k} \overline{F_{i k}} =\\
\sum_{k} |F_{ik}|^2 D_{k k} 
=\frac{1}{N} \tr D. 
\end{split}
\end{equation}
Consider now any hermitian matrix $H$ and denote by $U$ the matrix of its eigenvectors. 
Thus taking a unitary matrix  $U_{\max}=F U_{min}$  we see that the transformed matrix 
\begin{equation}
\label{contr1}
A=U_{max}HU_{max}^{\dagger}=F U_{min} H U_{min}^{\dagger} F^{\dagger}
\end{equation}
is contradiagonal, as all its diagonal elements are equal, as stated in item b).

To prove item a) we note, that the sum  in (\ref{maxA}) is maximized, if
and only if the sum
\begin{equation}
\sum_{i} |(VHV^{\dagger})_{ii}|^2 
\end{equation}
is minimized, since the vector of diagonal elements majorizes the constant
vector of the same sum, and the sum of squares is a Schur-convex function we obtain
the result.

The last item c) is obtained from the definition of the function $f$
by computing the trace of $H^2$ and subtracting the sum of squared elements at the diagonal.

A generalization of this procedure allowing to find a basis in which Hermitian matrix
with spectrum $y$ has diagonal $x \prec y$ is  presented in Appendix~\ref{sec:robin-hood}.

\medskip 
\begin{proposition}
 \label{lemma:max-dist}
Consider a family of unitarily similar hermitian matrices
\begin{equation}
G = V D V^{\dagger}, 
\end{equation}
where $D$ is a given diagonal matrix, then the maximal Hilbert-Schmidt distance
between the orbit of the unitarily similar matrices to diagonal matrix $D$
optimized with respect to all permutation matrices is given by
\begin{equation}
\max_{V \in U(N)} \min_{P \in Perm}\! \| D - PVDV^{\dagger}P^T\|^2_{\mathrm{HS}}
=
2 \left( \tr D^2 - \frac{(\tr D)^2}{N}\right),
\end{equation}
and for the optimal matrix $V_{\mathrm{opt}}$ one can take any complex Hadamard
matrix $F$.
\end{proposition}
\medskip 

Consider an arbitrary  hermitian matrix $H$, such that $H= U D U^\dagger$, so
its diagonalization corresponds to transforming the basis by the matrix $U$
consisting of eigenvectors. The above lemma explains why representing it in the
basis $W = F U ^\dagger$ can be called contra--diagonalization, as the matrix
$A = W H W^\dagger$ has all diagonal elements equal and is as far from the
diagonal matrix as possible.

%\subsection{$min-max$ definition for diagonal matrices}
%Let us consider a following problem, for a given diagonal matrix $D$ find a solution for
%\begin{equation}
%\max_{V \in U(N)} \min_{P \in Perm} \| D - PVDV^{\dagger}P^T\|^2_{\mathrm{HS}}.
%\end{equation}

\medskip 
\proof[Proof of Proposition~\ref{lemma:max-dist}]
To prove the lemma we write
\begin{equation}
\begin{split}
\max_{V \in U(N)} \min_{P \in Perm} \| D - PVDV^{\dagger}P^T\|^2_{\mathrm{HS}}
=\\
\max_{V \in U(N)} \min_{P \in Perm} 2 \tr D^2 - 2 \tr D P V D V^{\dagger}P^T.
\end{split}
\end{equation}
Next we note, that 
\begin{equation}
\max_{P \in Perm} \tr D P V D V^{\dagger}P^T = 
\max_{P \in Perm} \langle d | P q^{(V)}\rangle, 
\end{equation}
where $d$ is a diagonal of matrix $D$ and $q^{(V)}$ is a diagonal of matrix $ V
D V^{\dagger}$. It is easy to see that one obtain the maximum value if the
vectors are ordered in the same way, i.e.
\begin{equation}
\max_{P \in Perm} \langle d | P q^{(V)}\rangle = 
\langle d^{\downarrow} |  (q^{(V)})^{\downarrow}\rangle.
\end{equation}
To perform minimization over the set of unitary matrices $V\in U(N)$ we note
that the minimum value for the above  inner product is achieved, 
if vector $q$ is minimal in the majorization partial order, 
\begin{equation}
\min_{V \in U(N)}\langle d^{\downarrow} |  (q^{(V)})^{\downarrow}\rangle 
\leq \frac{\sum d_i}{N} \sum d_i = \frac{(\tr D)^2}{N}.
\end{equation}
The above minimum can be achieved if the unitary matrix $V$ is complex Hadamard 
for instance the Fourier matrix $F_N$.

To summarize the proof we write
\begin{equation}
\begin{split}
\max_{V \in U(N)} \min_{P \in Perm} \| D - PVDV^{\dagger}P^T\|^2_{\mathrm{HS}}
\\
= 2 \left( \tr D^2 - \frac{1}{N} (\tr D)^2\right),
\end{split}
\end{equation}
and every complex Hadamard matrix $V$ gives the maximum and in this case one can
take any permutation matrix $P$.
\hfill $\Box$

\subsection{Contradiagonal density matrices}

The statements on contra-diagonalization introduced above for arbitrary hermitian (or normal) matrices
can be now used for a positive definite density matrices $\rho^{\dagger}=\rho\ge 0$
normalized as $\tr \rho=1$. Thus a quantum state $\sigma$ of size $N$
will be called {\sl contradiagonal} if $\sigma_{ii}=1/N$ for $i=1,\dots, N$.

\begin{figure*}[htbp]
        \centerline{ \hbox{
        \epsfig{figure=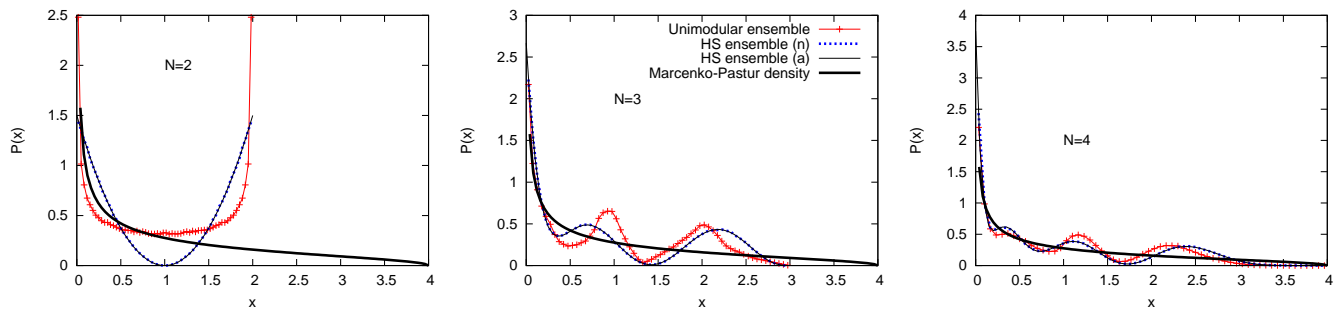,width=\linewidth} }}
     \caption{Distribution of squared singular values $P(x)$
    for random unimodular matrices 
    of order a) $N=2$, b) $N=3$ and c) $N=4$ (crosses).
    The case $N=2$ is described by the arcsin distribution supported in $[0,2]$,
   while the asymptotic behavior corresponds to the
     MP distribution represented by solid lines.
   The corresponding density of random states distributed according to the Hilbert-Schmidt measure
    is marked with dots for comparison, while the exact expressions are also plotted.}
     \label{fig_234}
\end{figure*}

Spectral density for the ensemble of contradiagonal states
obtained form a random unimodular matrix $A$ by Eq. (\ref{AAdag})
 is shown Fig. \ref{fig_234} for $N=2,3,4$.
 For $N=2$ a random contradiagonal density matrix takes
 the form  $ \sigma= \frac{1}{2}
   \left[   \begin{array}{c c}
            1     &  z  \\
     {\bar z}   & 1 
    \end{array}   \right] , $
 where $z=e^{i \psi_1} +e^{i \psi_2}$
and the phases $\psi_1$ and $\psi_2$ are random.
Thus the rescaled eigenvalues of $\sigma$ are distributed 
according to the arcsin law, 
$P_{\rm As}(x)=1/ \pi \sqrt{x(2-x)}$ for $x \in (0,2)$,
as shown in Fig.~\ref{fig_234}.

Note oscillations of the level density $P(x)$
present for $N \ge 3$. Observe that the conjectures above yield all
 the moments for any value of the matrix size $N$. 
It is then a classic moment problem to find the corresponding density.
 Curiously, there exists densities that have the exact same moments on different intervals and are 
found in \cite{ZPNC11} as a sequence of densities that converges to the MP distribution. 
We were however unable to solve this moment problem, to find the actual oscillatory one 
that is found for the unimodular ensemble. While the appearance
 of multiple densities with the same moments is known in the literature \cite{Korner}, this seems
 to be a curious case as the densities have support in $(0,1]$. The fact that the densities diverge at 
the origin makes the current moment problem not belong to the class of Haussdorf moment problems that treat
 compact intervals and absolutely continuous densities \cite{Korner}.

\medskip

The Schur--Horn theorem  states \cite{BZ06}
 that for any density matrix 
$\rho$ its diagonal is majorized by the spectrum,
${\rm diag}(\rho) \prec {\rm eig}(\rho)$.
The uniform vector $x_*=\{1/N, \dots, 1/N\}$
is majorized by any other probability vector.
This observation implies the following fact 

\medskip 
\begin{proposition}
 \label{prop3}
Let $\sigma$ denotes a contradiagonal state of order $N$,
so that $\sigma_{ii}=1/N$, and let $U$ be a unitary matrix of order $N$.
Then the following majorization relation holds
\begin{equation}
{\rm diag}(\sigma) \ \prec \ {\rm diag}(U\sigma U^{\dagger})
\ \prec \ {\rm eig}(\sigma) .
\label{prec2}
\end{equation}
\end{proposition}
\medskip 

The above result provides an additional argument in favor of 
usage of the notion of a contra--diagonal form of a matrix,
as  $\sigma$ is distinguished by the majorization order (\ref{prec2})
and is opposite to the diagonal form of a density matrix.

Let $\rho$ denote an arbitrary density matrix, $U_{\rm min}$ the matrix of its
eigenvectors and  $U_{\rm max}=F U_{\rm min}$ the matrix defining the bases in which
the state is contradiagonal. 
Then the entropy of the projective measurement of $\rho$
with respect to the basis $U_{\rm max}$ is maximal and equal to $\ln N$. 
Note that this basis
is hence dual to the eigenbasis of $\rho$ for which the entropy of the
projective measurement is minimal and equals to the von Neumann entropy
$S(\rho)$.
As each projective measurement induces the decoherence to the system
and copies the information on the eigenstates of the density matrix
to the environment, the information copied in the case of the measurement
in the contra-diagonal basis is the largest and reads $\ln N- S(\rho)$.
In other words, performing a coarse--graining map, $\rho \to \rho'={\rm diag}(\rho)$ 
on any pure state $\rho=|\psi\rangle\langle \psi|$, the exchange entropy \cite{Schum}
is the largest if the state is represented in the contra-diagonal basis.

Consider, for instance, a single--qubit pure state written its eigenbasis
as $H={\rm diag}(1,0)$. Making use of the real Hadamard matrix $F_2$
and putting it into 
Eq. (\ref{contr1}) one gets  the contradiagonal state $\sigma$ where
 \begin{equation}
F_2 = \frac{1}{\sqrt{2}}
   \left[   \begin{array}{c c}
            1     &  1  \\
            1   & -1 
    \end{array}   \right] 
{\ \ \rm and \ \ }
\sigma= \frac{1}{2}
   \left[   \begin{array}{c c}
            1     &  1  \\
            1   & 1 
    \end{array}   \right] .
\end{equation}
Choosing a complex Hadamard matrix $F_2'$ enphased with an
arbitrary complex phase $e^{\i \phi}$ we obtain a more general 
contradiagonal state ${\sigma}'$ with
 \begin{equation}
F_2' = \frac{1}{\sqrt{2}}
   \left[   \begin{array}{c c}
            e^{i \phi}     &  e^{i \phi}  \\
            1   & -1 
    \end{array}   \right],  
\ \ \
\sigma' = \frac{1}{2}
   \left[   \begin{array}{c c}
            1     &   e^{i \phi}  \\
     e^{-i \phi}  & 1 
    \end{array}   \right] .
\end{equation}
In higher dimensions, a density matrix $\vert \psi \rangle \langle \psi \vert$
corresponding to a pure state of size $N$
with spectrum ${\rm diag}(1,0, \dots, 0)$ transformed as in (\ref{contr1}) by the Fourier matrix $F_N$
leads to a flat contradiagonal state $\sigma$ with all elements equal, $\sigma_{ij}=1/N$.
Multiplying $F_N$ from left and right
by two arbitrary diagonal unitary matrices
one obtains an enphased, complex Hadamard $F_N'$ \cite{TZ06},
which leads to a more general form of a complex Hermitian contradiagonal state ${\sigma}'$
with all elements of the same modulus, $|{\sigma}'_{ij}|=1/N$,
and all diagonal elements equal, ${\sigma}'_{jj}=1/N$.

As stated in Proposition~\ref{lemma:anti} the sum of squared moduli of the off
diagonal elements is maximal if and only if the matrix is in its contra--diagonal form.
The above is equivalent to the fact, that the sum of squared diagonal elements
is minimal. On the other hand, since the geometric mean is a Schur concave
function, the product of diagonal elements of a semi positive matrix $H$ is
maximal if $H$ is contradiagonal. This fact was used in the
analysis of an entanglement measure called
\emph{collectibility}~\cite{rudnicki2012collectibility}.

\section{Concluding remarks}

In this work we showed that a generic random diagonal gate acting on a 
symmetric bipartite system is strongly non-local and is
amazingly efficient in generating quantum entanglement.
Its average Schmidt strength \cite{NDDGMOBHH02} is on average smaller than this
characteristic of a Haar random  unitary gate
by the factor of two, while the mean entangling powers 
\cite{ZZF00} for both ensembles are only marginally different.

Investigation of the ensemble of diagonal unitary gates of size $N^2$
leads to the unimodular ensemble of matrices of order $N$ with all entries
of the same modulus and independent random phases. We computed first moments of the distribution of the squared singular values of these matrices
and showed that it asymptotically converges to the universal Marchenko-Pastur form.
The moments have remarkable connections to combinatorial structures that have been recently studied. 
This allowed us to find the mean entanglement, or von Neumann entropy exactly for the ensemble of unimodular matrices.
However, for a finite $N$ we reported the differences with respect to the Hilbert-Schmidt ensemble, which produces, 
on average, less mixed states. 

Squared singular values of a matrix from the unimodular ensemble
after a suitable normalization coincide with the spectrum of 
a density matrix $\sigma$ of order $N$, such that all their diagonal elements are equal.
This form of a matrix is called contradiagonal, as it is shown to be opposite
to the diagonal form of a matrix concerning the majorization order, the norm of the off-diagonal elements
and the entropy of the projective measurement performed on a mixed state in such a basis.

In general, for any Hermitian matrix $H$ of order $N$
we have shown how to find a unitary 
matrix $U_{\rm max}$ which brings it to the contra--diagonal form, 
$H'=U_{\rm max}H U_{\rm max}^{\dagger}$ such that $H'_{jj}={\tr}H/N$.
This method based on diagonalization and complex Hadamard matrices \cite{TZ06} 
can be easily applied for any mixed state $\rho$ to transform it to its
contradiagonal form, distinguished form the perspective of
quantum information processing.

From a mathematical perspective the problem of 
constructing a Hermitian matrix with a given spectrum and prescribed diagonal entries 
was studied in \cite{Chu95,DHST05} and more recently in \cite{FMPS13}.
Although several general algorithms for this task were analyzed in these papers,
the limiting problem of a constant diagonal was not shown to be reducible
to the standard diagonalization procedure followed by a unitary transformation
with an arbitrary  complex Hadamard matrix.

\medskip

It is a pleasure to thank Sarika Jalan
for the invitation to the CNSD, at IIT Indore, where this work has been initiated.
We are thankful to Pawe{\l} Horodecki for 
stimulating discussions and fruitful remarks.
We acknowledge support by the grants number DEC-2011/02/A/ST1/00119 % NCN MAN_UJ
and DEC-2012/04/S/ST6/00400 (ZP) financed by Polish National Science Center.

\medskip

\appendix

\section{Reshuffling and operator entanglement entropy }

In this appendix a link between the operator Schmidt decomposition
and an the reshuffling of a matrix  \cite{ZB04} is reviewed. 

Consider a given unitary matrix $U$ of size $N^2\times N^2$.
It belongs to the composite Hilbert--Schmidt space ${\cal H}_{HS} \otimes {\cal
H}_{HS}$. 
Let us write down its
representation in a product basis in the space of matrices,
\begin{equation}
U  =  \sum_{m=1}^{N^2} \sum_{n=1}^{N^2} C_{mn}
 B_m \otimes B_n ,
 \label{Vmatrix}
\end{equation}
where $C_{mn}={\tr}( (B_m \otimes B_n)^{\dagger} U)$.

 The complex matrix $C$ of order $ N^2\times N^2$ 
need not be Hermitian nor normal. The usual Schmidt decomposition 
hold for this vector space and therefore
the Schmidt decomposition of $U$ given in Eq.(\ref{VSchmidt})
consists of $K$ terms entering with the weights $\sqrt{\Lambda_k}$ 
equal to the singular values of $C$.
%i.e. square roots of eigenvalues of the positive matrix $C^{\dagger}C$.

It will be convenient to work with the product bases in the HS space of matrices,
generated by the identity matrix, of size $N^2\times N^2$. 
Each of the $N^2$ basis matrices $B_n$ of size $N\times N$ 
has only a single non vanishing element equal to unity.
Let's denote $B_{k}=B^{m\mu}=|m\rangle\langle \mu |$, where
{\mbox{$k=N(m-1)+\mu$}}.

For this choice of the basis the matrix of the coefficients $C$ in  
Eq. (\ref{Vmatrix}) takes a particularly simple form, 
\begin{equation}
C_{\stackrel{\scriptstyle m \mu }{n
\nu}}= {\tr}(B^{m\mu}\otimes B^{n\nu})U= U_{\stackrel{ \scriptstyle mn}{\mu\nu}}.
\label{reschuff}
\end{equation}
Note that both matrices $U$ and $C$ consist of the same entries,
ordered in a different way.
This particular reordering of a matrix, called {\sl reshuffling} \cite{ZB04},
will be denoted as $U^R:=C$. In general the notion of reshuffling 
is well defined if a matrix $X$ acts on a composite Hilbert space, 
${\cal H}_M \otimes {\cal H}_N$. The symbol $U^R$ has a unique meaning if a
concrete decomposition of the total dimension, $L=MN$,  is specified. Similar
reorderings of matrices were considered by Hill et al. \cite{OH85,YH00} while
investigating CP maps and also  in \cite{Rud02,CW03,HHH02} to
analyze separability of mixed quantum states and in \cite{Shudd13} to generate local unitary invariants.
 This operation in these latter  contexts is also referred to as {\sl realignment}.

To get a better feeling of the transformation of reshuffling
 observe that reshaping each row of an arbitrary matrix 
$X$ of length $N^2$  into a submatrix of size $N$ and placing it according to the
lexicographical order block after block produces the reshuffled matrix $X^R$
as defined in (\ref{reschuff}).
Let us illustrate this procedure for the simplest case $N=2$, in which any
row of the matrix $X$ is reshaped into a $2 \times 2$ matrix
\begin{equation}
  C_{kj}=X_{kj}^R :=\left[
    \begin{array}{c|c}
      {\bf {X_{11}\ \ X_{12}}}  & X_{21}{\rm ~ ~ ~ }X_{22} \\
      X_{13} {\rm ~ ~ ~ }X_{14} & {\bf X_{23} \ \ X_{24}} \\
      \hline
      {\bf X_{31}\ \ X_{32}} & X_{41} {\rm ~ ~ ~ }X_{42} \\
      X_{33}{\rm ~ ~ ~ }X_{34} & {\bf X_{43}\ \ X_{44} }
    \end{array}
  \right] .
  \label{reshuf1}
\end{equation}
 It is easy to see that $(X^{R})^{R}=X$. In general,
$N^{3}$ elements of $X$ do not change their position during the operation of
reshuffling, these are typeset in bold in Eq.~(\ref{reshuf1}).
 the other $N^{4}-N^{3}$ elements do.
It is worth to emphasize that if a matrix $X$ is Hermitian 
the reshuffled matrix $X^R$ needs not to be Hermitian.

The Schmidt coefficients of $U$ are thus equal to  squared
singular values $\Lambda_i$ of the reshuffled matrix,
 $U^R$, equal to  the eigenvalues  of a positive matrix $H=(U^R)^{\dagger}U^R$.
%The initial basis is transformed by a local unitary
%transformation $W_a \otimes W_b$, where $W_a$ and $W_b$ are matrices of
%eigenvectors of matrices $(X^R)^{\dagger}X^R$ and $X^R(X^R)^{\dagger}$,
%respectively. 
The gate $U$ is local if and only if the rank $K$ of $H$ is equal to one, 
 so that the matrix can be factorized into a product form,
 $U=U_A \otimes U_B$.

The squared Hilbert-Schmidt norm of any unitary matrix of order $N^2$
is $||U||^2=N^2$,
which implies that $ \sum_{k=1}^{N^2} \Lambda_k=N^2$.
To characterize nonlocality of a gate $U$ one can then
use the normalized vector $\vec \lambda$ of the squared singular values,
$\lambda_k:=\Lambda_k/N^2$,
% lives in the ($N^2-1$) dimensional simplex and 
which may be interpreted as a probability vector of length $N^2$.

In general, the vector of the Schmidt coefficients of an
unitary matrix $U$ acting on a composite $N \times N$ system conveys
information concerning the non-local properties of $U$. To characterize
quantitatively the distribution of $\vec \lambda$ one uses the Shannon entropy,
\begin{equation}
  S(U):= S({\vec \lambda}) = - \sum_{k=1}^{N^2} \lambda_k \ln(\lambda_k)
  \label{shannon}
\end{equation}
called in this context {\sl entropy of entanglement} of $U$ \cite{Za01},
(or {\sl Schmidt strength} \cite{NDDGMOBHH02}), and the generalized,
R{\'e}nyi entropies
\begin{equation}
  S_q(U):= S_q({\vec \lambda})  =
-\frac{1}{1-q} \ln \Bigl[ \sum_{k=1}^{N^2} (\lambda_k)^q \Bigr],
  \label{Renyi}
\end{equation}
which tend to $S$ in the limit $q\to 1$. The entropy $S_0$, sometimes called {\sl
Hartley entropy}, is equal to $\ln K$, where $K$ denotes the number of positive
coefficients $\lambda_i$, and is called  {\sl  Schmidt rank} (or Schmidt number).
The second order R{\'e}nyi entropy $S_2$ is closely related to the linear entropy
$E(U)=1-\exp(-S_2)$ used by Zanardi in \cite{Za01}.

The generalized entropies $S_q$ are equal to zero 
if and only if the gate $U$ has a product structure, so it can be obtained
by performing local gates.
The upper bound, $S_q^{\rm max}=2\log N$
is achieved e.g. for the Fourier unitary matrix of size $N^2$ defined by
\begin{equation}
F_{kl}^{(N^2)}\ := \ \frac{1}{N}\exp\bigl(i2\pi kl/N^2\bigr)\ .
  \label{Four}
\end{equation}
 To show this fact it is sufficient to notice 
that the reshuffled matrix $F^R$ remains unitary,
 so all its singular values are equal to unity, hence the
Schmidt vector contains $N^2$ equal components and is maximally mixed.

\section{Hermitian matrices with prescribed spectrum} \label{sec:robin-hood}
We begin with a fact, known as a Horn lemma~\cite{BZ06}
\begin{lemma}
Assume that $x \prec y$ then there exist an orthostochastic matrix $\mathcal{O}$
such, that $x = \mathcal{O} y$.
\end{lemma} 
The matrix $\mathcal{O}$ is said to be orthostochastic, if there exist
an orthogonal matrix $W$, such that $\mathcal{O}_{ij} = W_{ij}^2$.

The above lemma in the case of bistochastic matrix instead of orthostochastic
is well known~\cite{hj2}, and sometimes used in a definition of majorization.

The Horn lemma allows us to formulate a simple lemma, which is a generalization of Proposition~\ref{lemma:anti}(b).
\begin{lemma}
Let $H$ be a Hermitian matrix with spectrum $y$ and let $x  \prec y$, then
there exist a unitary matrix $V$, such that the diagonal of $V H V^{\dagger}$
is given by $x$.
\end{lemma}
To prove it we assume, without loss of generality,
that $H$ is in its diagonal form with vector $y$ on diagonal,
and let $\mathcal{O}$
be an orthostochastic matrix such that $x = \mathcal{O} y$. By $W$ we denote
an orthogonal matrix such that $\mathcal{O}_{ij} = W_{ij}^2$. Now we write
\begin{equation}
(W H W^{T})_{ii} = \sum_{k} W_{ik} H_{kk} W_{ik} =\sum_{k} \mathcal{O}_{ik} y_{k} = x_i.
\end{equation}
In other words the unitary matrix $V$ which describes the unitary transformation
can be represented as a product of a unitary matrix $W$ appearing in the
Horn lemma and the matrix $U^{\dagger}$ containing eigenvectors of $H$.
In particular if $x$ is flat, i.e. all $x_i$ are equal, the matrix $W$ present
in the Horn lemma can be taken as a complex Hadamard matrix and item (b) in 
Proposition~\ref{lemma:anti} is recovered.

This allows us to obtain an alternative solution to the problem of constructing
Hermitian matrices with prescribed spectrum, studied in~\cite{Chu95,DHST05,FMPS13}.

\end{document}